\documentclass[twocolumn,superscriptaddress,letter]{revtex4}%
\usepackage{amssymb}
\usepackage{color}
\usepackage{graphicx}
\usepackage{dcolumn}
\usepackage{bm}
\usepackage[header,title,page,titletoc]{appendix}
\usepackage{amsmath}
\usepackage{amsfonts}%
\providecommand{\U}[1]{\protect \rule{.1in}{.1in}}

\begin{document}
\title{Theory of Ferrimagnetism in the Hubbard Model on Bipartite Lattices with
Spectrum Symmetry}
\author{Yang Xue}
\affiliation{Department of Physics, Beijing Normal University, Beijing, 100875, P. R. China}
\author{Jing He}
\affiliation{Department of Physics, Hebei Normal University, HeBei, 050024, P. R. China}
\author{Xing-Hai Zhang}
\affiliation{Department of Physics, Beijing Normal University, Beijing, 100875, P. R. China}
\author{Su-Peng Kou}
\thanks{Corresponding author}
\email{spkou@bnu.edu.cn}
\affiliation{Department of Physics, Beijing Normal University, Beijing, 100875, P. R. China}

\begin{abstract}
In this paper we developed theory of the ferrimagnetism in the Hubbard model
on bipartite lattices with spectrum symmetry. We then study the defect-induced
ferrimagnetic orders in three models and explored the universal features.

\end{abstract}
\maketitle

The magnetic orders are important magnetic properties for two dimensional (2D)
Hubbard model, to which researchers also have payed much attention. At half
filling case, the ground state is known to be the long-range (LR)
antiferromagnetic (AF) order. Nagaoka made a surprising discovery about the
induced ferromagnetic (FM) by a single hole for infinite coupling limit
$U\rightarrow \infty$\cite{na}. At heavily doping region a possible FM order
may exist. The starting point on the issue of \emph{ferrimagnetism} of the
Hubbard model is a theorem by Lieb\cite{lieb}. In this theorem, it is pointed
out that the total spin $S$ of the ground state for the Hubbard model on
bipartite lattice with particle-hole (PH) symmetry and real hopping parameters
is given by $S=|N_{\mathrm{A}}-N_{\mathrm{B}}|/2$ (we call it \emph{Lieb spin
moment}) where $N_{\mathrm{A}}$ and $N_{\mathrm{B}}$ are the numbers of
lattice sites on $\mathrm{A}$ and $\mathrm{B}$ sublattice, respectively. Then,
in Ref.\cite{shen}, it is pointed out that the ground state of the 2D
sublattice-unbalanced Hubbard model on bipartite lattice with the same
conditions (PH symmetry and real hopping parameters) is really the
ferrimagnetic (FR) order that possesses a finite total magnetic moment and the
LR AF order and obeys $m(0)\leq m(\mathcal{Q})$ (we call it Shen-Qiu-Tian
(SQT) inequality\cite{com1}) where $m(0)=%
{\displaystyle \sum \limits_{ij}}
\left \langle \hat{s}_{i}^{+}\hat{s}_{j}^{-}\right \rangle /N$ and
$m(\mathcal{Q})=%
{\displaystyle \sum \limits_{ij}}
(-1)^{i+j}\left \langle \hat{s}_{i}^{+}\hat{s}_{j}^{-}\right \rangle /N$
($\hat{s}_{i}=\frac{1}{2}\hat{c}_{i}^{\dagger}\sigma \hat{c}_{i},$ $N$\ is the
total number of the lattice sites). However, the FR order was seldom studied
\cite{shen,tian1,tian2,shen1} and the detailed properties of the FR order in
2D\ Hubbard model have not been explored.

\textit{The 2D Hubbard Model}: Our starting point is the following
Hamiltonian
\begin{equation}
H=-\sum \limits_{{i,j}}t_{ij}\left(  \hat{c}_{i}^{\dagger}\hat{c}%
_{j}+h.c.\right)  +U\sum \limits_{i}\hat{n}_{i\uparrow}\hat{n}_{i\downarrow
}-\mu \sum \limits_{i,\sigma}\hat{c}_{i\sigma}^{\dagger}\hat{c}_{i\sigma},
\end{equation}
where $\hat{c}_{i}=\left(  \hat{c}_{i},_{\uparrow},\hat{c}_{i},_{\downarrow
}\right)  ^{T}$ with $\hat{c}_{i,\sigma}$ representing the fermion
annihilation operator at site $i$. $\sigma=\uparrow,\downarrow$ denote spin
index. For this bipartite lattice, we have two sub-lattices, \textrm{A} and
\textrm{B}. $t_{ij}$ is the hopping amplitude. $U$ is the strength of the
repulsive interaction. $\mu$ is the chemical potential which is set to $\mu=0$
for the half filling case.

\textit{The spectrum symmetry}: Firstly, we defined the \emph{spectrum
symmetry}. For the Hamiltonian in Eq.(1) with the spectrum symmetry (S
symmetry), the denisty of state (DOS) $\rho(E)$ is always symmetric via $E$ as
$\rho(E)=\rho(-E)$. As a result, each energy level with positive energy $E$
must be paired with an energy level with negative energy $-E$. To make it
clearer, we define an operator of the S symmetry as $\mathcal{S}%
=\mathcal{P}\cdot \mathcal{D}$ where $\mathcal{P}$ is the PH transformation
operator $\mathcal{P}=\mathcal{R}\cdot \mathcal{K}$ introduced in
Ref.\cite{mic,he} and $\mathcal{R}$ is an operator that leads to $\hat{c}%
_{i}\leftrightarrow(-1)^{i}\hat{c}_{i}$, $\mathcal{K}$ is the complex
conjugate operator, $\mathcal{D}$ is a discrete transformation operator that
commutes with lattice translation operators $T_{x,y}$ as, $\left[
\mathcal{D}\text{, }T_{x}\right]  =\left[  \mathcal{D}\text{, }T_{y}\right]
=0$. For a Hamiltonian with the S symmetry, we have $H=-\mathcal{S}^{\dagger
}H\mathcal{S}$. Thus, each energy level $\left \vert \psi \right \rangle $ with
positive energy $E$ is paired with an energy level $\mathcal{S}\left \vert
\psi \right \rangle $ with negative energy $-E$. So, the PH symmetry is a
special case of the S symmetry and for the case of $\mathcal{D}=1,$ the S
symmetry is reduced into the PH symmetry.

\textit{Vacancy-induced zero-modes}: We then consider the case of free
Hamiltonian ($U=0$) with lattice-defect - the vacancy by adding a potential on
given lattice site $R$, $H_{R}=-\sum \limits_{{i,j}}t_{ij}\left(  \hat{c}%
_{i}^{\dagger}\hat{c}_{j}+h.c.\right)  +V_{R}\hat{c}_{R}^{\dagger}\hat{c}_{R}%
$. In the unitary limit, the lattice-defect becomes a missing lattice site
that is just a vacancy, on which we have an infinite on-site potential, i.e.,
$V_{R}\rightarrow \infty$. It is pointed out that the vacancy doesn't break the
S symmetry ($H_{R}=-\mathcal{S}^{\dagger}H_{R}\mathcal{S}$). Due to the S
symmetry, there exists a zero energy state (the so-called zero-mode)
$\left \vert \psi_{0}\right \rangle $ when we add a vacancy on $\mathrm{A}%
$\textrm{ }sublattice. For the zero-mode (ZM) $\left \vert \psi_{0}%
\right \rangle ,$ we have $\left \vert \psi_{0}\right \rangle =\mathcal{S}%
\left \vert \psi_{0}\right \rangle $. We denote the wave-function of the ZM by
$\psi_{0}(r_{i}-R)$ where $R$ is the position of the vacancy. When there
exists $n$-vacancy on $\mathrm{A}$\textrm{ }sublattice, $n=\left \vert
N_{\mathrm{A}}-N_{\mathrm{B}}\right \vert $ zero energy modes will necessarily
appear. In addition, for the case with nearest neighbor hopping $t_{ij}%
=t_{\left \langle ij\right \rangle }$ ($t_{\left \langle ij\right \rangle }$ is
the real or complex nearest neighbor (NN) hopping parameter), these
vacancy-induced (VI) zero-modes (ZMs) localize only on the $\mathrm{B}$
sublattice and are orthotropic each other.

For some models with S symmetry, the DOS may be finite at the Fermi level.
Except for the VI ZMs, there may exist additional zero energy states
$\left \vert \psi \right \rangle \neq \mathcal{S}\left \vert \psi \right \rangle $.
However, the additional zero energy states are not protected by symmetry and
are fragile against perturbations. On the contrary, since the VI ZMs are
protected by S symmetry ($\left \vert \psi_{0}\right \rangle =\mathcal{S}%
\left \vert \psi_{0}\right \rangle $), they are fixed precisely at the Fermi
level, the interaction term is highly relevant. In particular, arbitrary small
(repulsive) interaction will drive the spin moments of the VI ZMs into an FM
ordered state. Consequently, the ground state of the original Hamiltonian
turns into a LR FR order.

\textit{Ferrimagnetism}: Let us show the universal features of the FR order in
the Hubbard model in the small $U$ limit.

For a system with $L_{x}\times L_{y}$ vacancy-lattice ($L_{x}$ along $x$
direction and $L_{y}$ along $y$ direction), in a unit cell (UC) there are
$L_{x}\times L_{y}/a^{2}-1$ lattice sites. In general, we have $L_{x}\times
L_{y}/a^{2}-1$ magnetic order parameters $M_{i}=\langle \hat{c}_{i\uparrow
}^{\dag}\hat{c}_{i\uparrow}-\hat{c}_{i\downarrow}^{\dag}\hat{c}_{i\downarrow
}\rangle/2$ to denote the local magnetizations on the lattice sites in a UC.
To simply characterize the FR order, we introduce two order parameters, the
total FM moment in a unit cell $\mathcal{M}=\sum_{i\in \text{unit-cell}}%
\langle \hat{c}_{i\uparrow}^{\dag}\hat{c}_{i\uparrow}-\hat{c}_{i\downarrow
}^{\dag}\hat{c}_{i\downarrow}\rangle$ and the total AF moment in a unit cell
$\mathcal{N}=\sum_{i\in \text{unit-cell}}(-1)^{i}\langle \hat{c}_{i\uparrow
}^{\dag}\hat{c}_{i\uparrow}-\hat{c}_{i\downarrow}^{\dag}\hat{c}_{i\downarrow
}\rangle$, respectively. One can see that $m(0)=%
{\displaystyle \sum \limits_{ij}}
\left \langle \hat{s}_{i}^{+}\hat{s}_{j}^{-}\right \rangle /N=%
{\displaystyle \sum \limits_{i}}
\left \langle \hat{c}_{i\uparrow}^{\dag}\hat{c}_{i\uparrow}-\hat{c}%
_{i\downarrow}^{\dag}\hat{c}_{i\downarrow}\right \rangle /N=\mathcal{M}%
/(L_{x}\times L_{y}/a^{2}-1)$ and $m(\mathcal{Q})=\mathcal{N}/(L_{x}\times
L_{y}/a^{2}-1)$. We then define the spin operator of the VI ZM as
$\mathbf{\hat{S}}_{R}(i)=\frac{1}{2}\hat{\psi}_{0}^{\dagger}(r_{i}%
-R)\mathbf{\sigma}\hat{\psi}_{0}(r_{i}-R)$ where $\hat{\psi}_{0}=(\hat
{c}_{0,\uparrow},\hat{c}_{0,\downarrow})^{T}\psi_{0}$ and $\hat{c}_{0,\sigma}$
is the\ particle annihilation operator of the zero-mode with spin $\sigma$.
When $\left \langle \hat{S}_{R}^{z}\right \rangle \neq0,$ the FR\ order is a
direct physics consequence of the FM\ order of the spin moments of the
different ZMs. The total FM moment in a UC is given by $\mathcal{M}%
=2\sum_{i\in \text{unit-cell}}M_{i}=2%
{\displaystyle \sum \limits_{R}}
\left \langle \hat{S}_{R}^{z}\right \rangle ,$ and the total AF moment in a UC
is $\mathcal{N}=2\sum_{i\in \text{unit-cell}}(-1)^{i}M_{i}=2%
{\displaystyle \sum \limits_{R}}
\left \langle (-1)^{i}\hat{S}_{R}^{z}\right \rangle .$

From above discussion, we already know that there exists a ZM for each
vacancy. In the small $U$ limit, $U\rightarrow0$, the low energy physics is
dominated by these ZMs. Because the ZMs induced by two different vacancies\ at
$R$ and $R^{\prime}$ are orthotropic each other, when considering the on-site
interaction there exists the Hund rule's coupling as $-J_{\mathrm{eff}%
}(R,R^{\prime})\mathbf{\hat{S}}_{R}\cdot \mathbf{\hat{S}}_{R^{\prime}}$ where
$J_{\mathrm{eff}}(R,R^{\prime})>0\ $is the effective FM spin coupling
constant. Thus, the low energy effective Hamiltonian becomes $H_{\mathrm{eff}%
}=-%
{\displaystyle \sum \limits_{R,R^{\prime}}}
J_{\mathrm{eff}}(R,R^{\prime})\mathbf{\hat{S}}_{R}\cdot \mathbf{\hat{S}%
}_{R^{\prime}},$ of which the ground state is a long range FM order denoted by
$\left \langle \hat{S}_{R}^{z}\right \rangle =\frac{1}{2}\left \langle \hat{\psi
}_{0,R\uparrow}^{\dagger}\hat{\psi}_{0,R\uparrow}-\hat{\psi}_{0,R\downarrow
}^{\dagger}\hat{\psi}_{0,R\downarrow}\right \rangle =\frac{1}{2}$ or
$\mathcal{M}=1$. As a result, the total spin moment of the ground state must
be the total number of spin moments of the VI ZMs $S=\frac{1}{2}%
{\displaystyle \sum}
\mathcal{M}$ that is just the Lieb spin moment $S=|N_{\mathrm{A}%
}-N_{\mathrm{B}}|/2$. The local magnetizations are given by $M_{i}\rightarrow%
{\displaystyle \sum \limits_{R}}
\left \langle \hat{S}_{R}^{z}(i)\right \rangle .$

For the case with only NN hopping $t_{ij}=t_{\left \langle ij\right \rangle }$
in small $U$ limit, it is obvious that $M_{i\in \mathrm{A}}=0.$ Thus, we have
$\mathcal{N}=\mathcal{M}=1.$ $\mathcal{M}=\mathcal{N}$ means the ground state
is an FM-AF-balanced FR order. On the contrary, in the strong coupling limit,
$U\rightarrow \infty,$ the low energy effective Hamiltonian turns into the
(un-frustrated) Heisenberg model with vacancy-lattice as $H_{\mathrm{eff}}=J%
{\displaystyle \sum \limits_{\left \langle ij\right \rangle }}
\mathbf{\hat{s}}_{i}\cdot \mathbf{\hat{s}}_{j}$ ($J\rightarrow \frac{4t^{2}}{U}%
$). The ground state is characterized by the AF ordered staggered
magnetization, $M=2(-1)^{i}M_{i}\rightarrow1/2$. So, we have $\mathcal{N}%
=2(L_{x}\times L_{y}/a^{2}-1)M\rightarrow L_{x}\times L_{y}/a^{2}-1$ and
$\mathcal{M}=1$. The LR FR order is really a defect-diluted AF
order\cite{lieb1,tian1}.

For case with both NN hopping $t_{ij}=t_{\left \langle ij\right \rangle }$ and
next nearest neighbor (NNN) hopping, $t_{ij}=t_{\left \langle \left \langle
ij\right \rangle \right \rangle }\neq0$, the situation becomes complex. The VI
ZM still exists due to S symmetry. However, the wave-functions of the ZM may
distribute on both sublattices. As a result, in small $U$ limit, we have
$\mathcal{M}=1$ and the total FM moment is also Lieb spin moment $|N_{A}%
-N_{B}|/2$. In the large $U$ limit, the low energy effective Hamiltonian turns
into the frustrated Heisenberg model, of which the ground state may be not an AF\ order.

\textit{Example 1 - the Hubbard model on square lattice}: For the Hubbard
model on square lattice, the hopping parameters are $t_{\left \langle
ij\right \rangle }=t$. For this model, the operator of the S symmetry is
$\mathcal{S}=\mathcal{P}$. The S symmetry protected VI ZM is an extended
state, of which the wave-function $\Psi_{0}$ can be naturally be $\psi
_{0,i\in \mathrm{A}}=0,$ $\psi_{0,i\in \mathrm{B}}=\frac{1}{N_{B}}$. See part of
the particle density distribution of this ZM on a $60a\times60a$ square
lattice in Fig.1(a).

\begin{figure}[ptb]
\includegraphics* [width=0.5\textwidth]{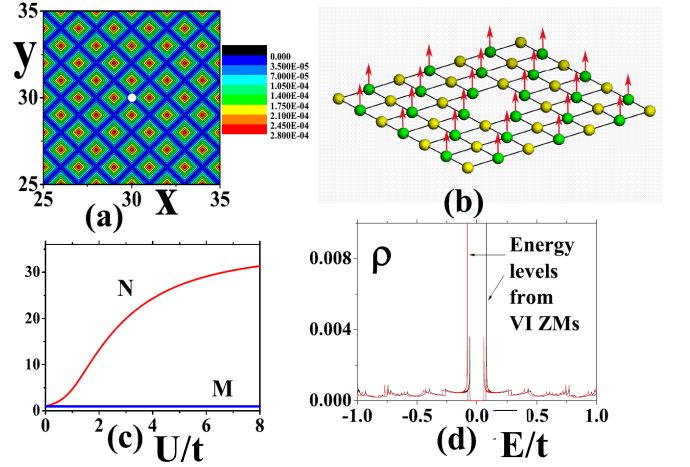}\caption{(Color online) (a)
Part of particle density distribution of VI ZM on a $60a\times60a$ square
lattice (a vacancy at the white spot); (b) The illustration of the uniform
FM-AF-balanced ferrimagnetic order for the Hubbard model on square lattice and
$U=0.01t$; (c) The two order parameters; (d) The density of state for case of
Hubbard model on square lattice and $U=t$.}\end{figure}

To check the validity of above discussion, we use the mean field approach to
study the Hubbard model on square lattice with a vacancy-lattice. The lattice
constant of the vacancy-lattice is set to be $d=6a.$ We choose $6a\times6a$
sites to be a UC ($6a$ along $x$ direction and $6a$ along $y$ direction). To
search the ground state with the lowest energy, we need to solve
$6\times6-1=35$ order parameters that denote the local magnetizations
$M_{i}=\langle \hat{c}_{i\uparrow}^{\dag}\hat{c}_{i\uparrow}-\hat
{c}_{i\downarrow}^{\dag}\hat{c}_{i\downarrow}\rangle/2$ on $35$ lattice sites
in a UC.

From the mean field calculation, we find that for the weak coupling limit,
$U\rightarrow0,$ the ground state is a \emph{uniform} Ferrimagnetic order,
$M_{i\in A}=0,$ $M_{i\in B}=\frac{1}{6\times6-1}=\frac{1}{35}$. The FR ordered
state is illustrated in Fig.1(b). From Fig.1(c) one can see that the total FM
moment in a UC is indeed a constant, $\mathcal{M}\equiv1$, which is consistent
to the prediction of Lieb spin moment $|N_{A}-N_{B}|/2$. On the other hand,
the total AF moment in a UC is also $\mathcal{N}=1$. The FM-AF-balance
character ($\mathcal{M}=\mathcal{N}$ or $m(0)=m(\mathcal{Q})$) comes from the
fact that the VI ZMs only distribute on one sublattice. With the increasing
the interaction strength, the average magnetizations on the sites of
\textrm{B} sublattice become finite values with an opposite polarization to
those on the sites of \textrm{A} sublattice $M_{i\in \mathrm{A}}$ :
$\frac{M_{i\in \mathrm{A}}}{\left \vert M_{i\in \mathrm{A}}\right \vert }%
=-\frac{M_{i\in \mathrm{B}}}{\left \vert M_{i\in \mathrm{B}}\right \vert }\neq0$.
Because the amplitudes of $M_{i}$ on all lattice sites increase, we have an
AF-dominated FR order with $\mathcal{M}<\mathcal{N}$ (or $m(0)<m(\mathcal{Q}%
)$). Now, the SQT inequality $m(0)\leq m(\mathcal{Q})$ is satisfied. In the
large $U$ limit, we have a saturated value, $\mathcal{N}\rightarrow35$ but
$\mathcal{M}\equiv1$. Fig.1(d) shows the DOS\ of the FR order, of which there
exists an energy gap. Near the gap, the DOS is enhanced due to the VI ZMs.

\textit{Example 2 - the staggered-flux Hubbard model on square lattice}:
Recently, people had realized the photon-assisted tunneling on optical lattice
and then generated a large effective (staggered) magnetic flux on optical
lattice\cite{ai,ai1,fl}. When two-component fermions with repulsive
interaction are put into such optical lattice, one can get an effective
staggered-flux Hubbard model. These progresses may provide new research
platform to learn the ferrimagnetism. It is easy to change the potential
barrier by varying the laser intensities to tune the Hamiltonian parameters
including the hopping strength ($t$-term), the staggered flux ($\phi$) and the
particle interaction ($U$-term). For this reason, we take the Hubbard model on
square lattice with staggered-flux (SF) as the second example, where
$t_{i,i\pm \delta_{y}}=e^{i\pm \phi}t$ for $i\in \mathrm{A}$, $t_{i,i\pm
\delta_{x}}=t$ for $i\in \mathrm{A}$ and $t_{i,i\pm \delta_{y}}=t_{i,i\pm
\delta_{x}}=t$ for $i\in \mathrm{B}$. See the illustration in Fig.2(a).\ In
particular, we only consider the NN hoppings. For this model, the operator of
the S symmetry is $\mathcal{S}_{T}=\mathcal{P}\cdot \mathcal{T}$ where
$\mathcal{T}$ is the time-reversal transformation operator which commutes with
lattice translation operators $\left[  \mathcal{T}\text{, }T_{x}\right]
=\left[  \mathcal{T}\text{, }T_{y}\right]  =0$.

After diagonalization of the Hamiltonian in momentum space, the energy spectra
are obtained as $E_{\pm}=-t\left[  \cos(k_{y}-\phi)+\cos k_{y}\right]
\pm \sqrt{t^{2}\left[  \cos(k_{y}-\phi)-\cos k_{y}\right]  ^{2}+4t^{2}\cos
^{2}k_{x}}.$ For the case of $\phi=0$, the SF disappears and we get a uniform
Hubbard model on square lattice. In the BZ $k_{x}\in(-\pi,\pi)$, $k_{y}%
\in(-\pi,\pi)$, the energy S turns into $E=-2t(\cos k_{x}+\cos k_{y}).$ Now
the Fermi surface at half filling has perfect nesting condition. Away from
this case, $\phi \neq0$, the BZ is reduced into a half one. The system becomes
a semi-metal and also has the perfect nesting condition. For the $+$ band,
there exists a hole pocket; For the $-$ band, there exists an electron pocket.
See the illustration in Fig.2(b). The density of state (DOS) $\rho(E)$ near
Fermi surface is reduced with increasing $\phi$. For the case of $\pi$-flux,
the energy S turns into $E_{\pm}=\pm2t\sqrt{\cos^{2}k_{y}+\cos^{2}k_{x}}$ and
the electron pocket and hole pocket shrinks into two Dirac nodes at
$\mathbf{K}_{1}=(\pi/2,\pi/2)$ and $\mathbf{K}_{2}=(\pi/2,-\pi/2)$.

\begin{figure}[ptb]
\includegraphics* [width=0.5\textwidth]{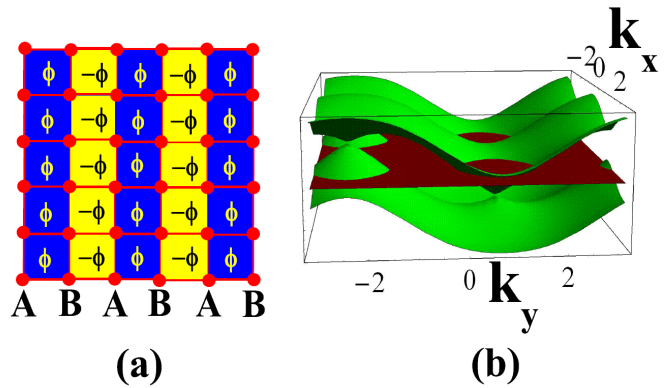}\caption{(Color online) (a)
The illustration of the staggered-flux lattice; (b) The dispersion of the
$\pi/2$-flux case: there are a hole pocket and an electron pocket in a reduced
BZ. The red plane denotes the position of chemical potential.}%
\end{figure}

Because the SF\ Hubbard model on bipartite lattices at half-filling is
unstable against antiferromagnetic (AF) instability, the ground state becomes
an insulator with AF order for the case of finite $U$. Such AF order is
described by the following mean field ansatz $\langle \hat{c}_{i,\sigma}^{\dag
}\hat{c}_{i,\sigma}\rangle=\frac{1}{2}\left(  1+(-1)^{i}\sigma M\right)  $
where $M$ is the staggered magnetization. For the cases of spin up and spin
down, we have $\sigma=+1$ and $\sigma=-1$, respectively. Then in the mean
field theory, by minimizing the ground state energy in the reduced Brillouin
zone, we could solve the staggered magnetization. Due to the perfect nesting
condition, the arbitrary small interaction term leads to an AF
spin-density-wave (SDW) order and then the BZ of $\phi \neq0,$ $\pi$ case is
reduced into a quarter one. For the $\pi$-flux case, the DOS near Fermi
surface is zero and the critical point between the semi-metal and AF insulator
is about $U_{c}=3.11t$\cite{hsu,sun}.

For the case of $\phi \neq0,$ the VI ZM is a quasi-localized state. See the
particle density distribution of this ZM for the case of $\phi=\pi/2$ in
Fig.3(a). The VI ZMs are anisotropic due to the rotation-symmetry breaking of
the original Hamiltonian. In the continuum limit, for the case of $\phi=\pi,$
the wave function of VI ZM distributes on $\mathrm{B}$ sublattice that has a
simple form of $\psi_{0}(x,y)\simeq \frac{e^{i\mathbf{K}_{1}.\mathbf{r}}}%
{x+iy}+\frac{e^{i\mathbf{K}_{2}.\mathbf{r}}}{x-iy}$\cite{per}. The amplitude
of this state decays with the distance to the vacancy as $1/r$. It is needed
to point out that these VI ZMs are all protected by the S symmetry.

\begin{figure}[ptb]
\includegraphics* [width=0.5\textwidth]{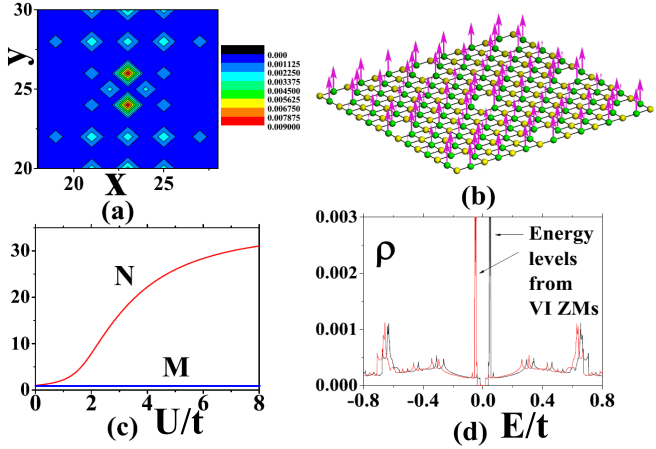}\caption{(Color online) (a)
Part of particle density distribution of VI ZM on a $60a\times60a$ square
lattice for $\phi=\pi/2$ and $U=0$; (b) The illustration of the cluster
AF-balanced ferrimagnetic order for $\phi=\pi/2$ and $U=0.001t$; (c) The two
order parameters for $\phi=\pi/2$; (d) The density of state for $\phi=\pi/2$
and $U=t$. }%
\end{figure}

We use the mean field approach to study the Hubbard model on SF square lattice
with a $6a\times6a$ vacancy-lattice. We focus on the case of $\phi=\pi/2$. Now
the ground state is a \emph{cluster} Ferrimagnetic order for weak coupling
case. The word "cluster" means that the local magnetic order parameters
$M_{i}$ is larger near the vacancy but smaller far from it. See the
illustration Fig.3(b). This "cluster" behavior obviously is a physical
consequence of the quasi-localized VI ZMs. From Fig.3(c) one can also see that
$\mathcal{N}\rightarrow1$, $\mathcal{M}\equiv1$ ($U\rightarrow0$) and
$\mathcal{N}\rightarrow35$, $\mathcal{M}\equiv1$ ($U\rightarrow \infty$). From
the DOS of the system (Fig.3(d)), one can see that there exists energy gap of
the mid-gap states (zero-modes) that dominate the low energy physics. All
these features indicate an AF-dominated FR\ order from the FM order of spin
moments of the quasi-localized ZMs.

From the mean field calculation, for the case of $\phi=\pi,$ we find that the
quantum phase transition between the metallic (or semi-metallic) states and
the magnetic orders shifts from $U=U_{c}$ to $U=0$. Arbitrary interaction
drives the system into a long range FR order, which is obviously induced by
the vacancy-lattice. At $U\sim U_{c}$, there is no true phase transition,
instead, a crossover occurs. For $U<U_{c},$ the ground state can be regarded
as an FR order from the FM\ ordered spin moments of ZMs. Now, a tiny energy
gap opens which is due to the spin polarized effect of the ZMs. On the other
hand, for $U>U_{c},$ the ground state can be regarded as the defect-diluted AF
order, and a big energy gap opens which is just AF-Mott gap for the double
occupied particles on one site.

\begin{figure}[ptb]
\includegraphics* [width=0.5\textwidth]{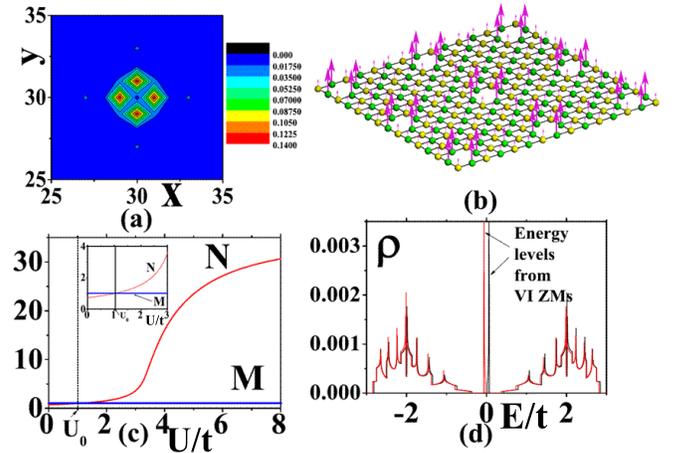}\caption{(Color online) (a)
Part of particle density distribution of VI ZM for the interacting spinful
Haldane model on $60a\times60a$ square lattice at $\phi=\pi,$ $t^{\prime
}=0.1,$ and $U=0$; (b) The illustration of the cluster FM-balanced
ferrimagnetic order for the interacting spinful Haldane model on square
lattice at $\phi=\pi,$ $t^{\prime}=0.1$ and $U=0.001t$; (c) The two order
parameters of the case of $\phi=\pi,$ $t^{\prime}=0.1$; (d) The density of
state of the interacting spinful Haldane model on square lattice at $\phi
=\pi,$ $t^{\prime}=0.1$ and $U=t$. }%
\end{figure}

\textit{Example 3 - the spinful Haldane model on square lattice}: An
interesting issue is the FR order for the case from a topological insulator
with NNN hoppings. Now we consider the spinful Haldane model on square
lattice, of which the Hamiltonian is given by $H_{T}=H(\phi=\pi)+H(t^{\prime
})$ where $H(\phi=\pi)$ is the Hamiltonian of the SF Hubbard model on square
lattice and $t^{\prime}$ is the NNN hopping. The Hamiltonian has PH symmetry
from $H_{T}=-\mathcal{P}^{\dagger}H_{T}\mathcal{P}$. In addition, the free
Hamiltonian is a topological Chern insulator.

We numerically calculated the free Hamiltonian with a vacancy, and found a (S
or PH symmetry protected) VI ZM on a $60a\times60a$ lattice\cite{he}. Fig.4(a)
shows the particle density of the ZMs, localized around the defect center
within a length-scale of $\sim(\Delta E)^{-1}$, where $\Delta E$ is the energy
gap of the vacancy-free case. In particular, the wave-function of the ZM
distribute not only on \textrm{B} sublattice but also on \textrm{A} sublattice.

We then use the mean field approach to study the FR order. From the results
given in Fig.4(b), we find that the ground state is also a cluster FR. In the
small $U$ limit, the average magnetizations on the sites of \textrm{A}
sublattice become \emph{finite} as $\frac{M_{i\in \mathrm{A}}}{\left \vert
M_{i\in \mathrm{A}}\right \vert }=\frac{M_{i\in \mathrm{B}}}{\left \vert
M_{i\in \mathrm{B}}\right \vert }\neq0$. Now, the magnetic order in \textrm{A}
sublattice has the same polarized direction to that of \textrm{B} sublattice
but the value of it is much smaller to that of \textrm{B} sublattice. The
total spin $S$ of the ground state still obeys the prediction of Lieb spin
moment as $|N_{\mathrm{A}}-N_{\mathrm{B}}|/2$. However, we have $\mathcal{M}%
>\mathcal{N}$ that mean $m(0)>m(\mathcal{Q})$. The ground state is an
FM-dominated FR order and the SQT inequality is violent. We conclude that the
violence of SQT inequality for this case is due to the NNN hoppings.

When we increase the interaction strength, $m(0)=m(\mathcal{Q})$ at $U_{0}$.
When we further increase the interaction strength, the ground state turns into
an AF-dominated FR order with $m(0)<m(\mathcal{Q})$. For larger interaction
strength, a topological quantum phase transition occurs. The energy gap closes
and opens again. The system then has no nontrivial topological properties and
becomes a defect-diluted AF order.

\textit{Conclusion}: In this paper, we developed a universal formula of the
ferrimagnetism in the Hubbard model beyond Lieb's theorem by taking into
account for the S symmetry with larger universality than traditional PH
symmetry. Then, by taking three models as examples, we study the
defect-induced FR orders that emerge from three typical fermionic systems -
metal, semi-metal, (Chern) insulator.\ We found that there may exist various
FR\ orders (uniform FM-AF-balanced FR order, uniform AF-dominated FR order,
cluster FM-AF-balanced FR order, cluster AF-dominated FR order, cluster
FM-dominated FR order...). The total spin of all these FR orders is equal to
the Lieb spin moment ($|N_{\mathrm{A}}-N_{\mathrm{B}}|/2$). From the common
feature of these FR orders, we conjecture that it is the S symmetry protected
VI ZMs ($\left \vert \psi_{0}\right \rangle =\mathcal{S}\left \vert \psi
_{0}\right \rangle $) that dominate the low energy physics of the system in
small $U$ limit. However, we found that the SQT inequality ($m(0)\leq
m(\mathcal{Q})$) is valid for the model with only the NN hoppings and can be
violent by the NNN hoppings. In addition, we may point out that the formula
can be straightforwardly applied to other Hubbard models on bipartite lattices
with S symmetry.

\begin{center}
{\textbf{* * *}}
\end{center}

This work is supported by National Basic Research Program of China (973
Program) under the grant No. 2011CB921803, 2012CB921704 and NSFC Grant No.
11174035.


\bigskip

\begin{thebibliography}{99}                                                                                               %


\bibitem {na}Y. Nagaoka, Phys. Rev. \textbf{147}, 392-405 (1966).

\bibitem {lieb}E. H. Lieb, Phys. Rev. Lett. \textbf{62}, 1201-1204 (1989),
[Errata \textbf{62}, 1927 (1989)].

\bibitem {shen}S. Q. Shen, Z. M. Qiu and G. S. Tian, Phys. Rev. Lett.
\textbf{72}, 1280 (1994).

\bibitem {com1}After a (partial) particle-hole transformation, the repulsive
Hubbard model can be transformed into an attractive Hubbard model, of which
the ground state always has singlet superconducting (SC) pairing. This singlet
SC just corresponds to the ground state with a total spin $|N_{\mathrm{A}%
}-N_{\mathrm{B}}|/2$\cite{lieb,tian,shen2}. Then, the SQT inequality
corresponds to non-negative off-diagonal long-range SC pairing of the
attractive Hubbard model in earlier paper\cite{tian}.

\bibitem {tian1}G. S. Tian, J. Phys. A: Math. Gen. \textbf{27} 2305 (1994).

\bibitem {tina3}G. S. Tian, Phys. Rev. \textbf{B 50,} 6246 (1994).

\bibitem {tian2}G. S. Tian and T. H. Lin, Phys. Rev. \textbf{B 53}, 8196 (1996).

\bibitem {shen1}S. Q. Shen, Int. J. Mod. Phys. \textbf{B 12}, 709 (1998).

\bibitem {mic}C. N. Yang and S. C. Zhang, Mod. Phys. Lett. \textbf{B 4,} 759 (1990).

\bibitem {he}J. He, et al, Phys. Rev. \textbf{B 87}, 075126 (2013).

\bibitem {ai}M. Aidelsburger, et al, Phys. Rev. Lett. \textbf{107,} 255301 (2011).

\bibitem {ai1}M. Aidelsburger, et al, Phys. Rev. Lett. \textbf{111}, 185301 (2013).

\bibitem {fl}H. Miyake, et al, Phys. Rev. Lett. \textbf{111}, 185302 (2013).

\bibitem {hsu}T. C. Hsu, Phys. Rev. B. \textbf{41}, 11379 (1990).

\bibitem {sun}G. Y. Sun and S. P. Kou, Europhys. Lett. \textbf{87,} 67002 (2009).

\bibitem {per}V. M. Pereira, et.al, Phys. Rev. Lett. \textbf{96,} 036801 (2006).

\bibitem {lieb1}E. H. Lieb, D. C. Mattis, J. Math. Phys. \textbf{3,} 749 (1962).

\bibitem {tian}G. S. Tian, Phys. Rev. \textbf{B 45}, 3145 (1992).

\bibitem {shen2}S. Q. Shen and Z. M. Qiu, Phys. Rev. Lett. \textbf{71}, 4238 (1993).
\end{thebibliography}
\end{document}